\documentclass {aa}
\usepackage{natbib}
\usepackage{amssymb}
\usepackage{amsmath}
\usepackage{graphicx}

\def\CGPLUS{{CG$^{+}$}}

\def\aap{A\& A}

\def\apj{ApJ}
\def\aj{AJ}

\def\AU{\mathrm{AU}}

\def\comma{\,,}
\def\fullstop{\,.}
\def\thttle{An analysis of two-layer models for circumstellar disks}
\begin{document}
\title{\thttle}
\author{C.P.~Dullemond and A.~Natta}
\authorrunning{Dullemond \& Natta}
\titlerunning{\thttle} 
\institute{Max Planck Institut f\"ur Astrophysik, Karl
Postfach 1317, D--85741 Garching, Germany; e--mail:
dullemon@mpa-garching.mpg.de
\and Osservatorio Astrofisico di Arcetri,
Largo E.~Fermi 5, 50125 Firenze, Italy}
\date{DRAFT, \today}

\abstract{The two-layer disk models of Chiang \& Goldreich (1997, henceforth
CG) and its derivatives are popular among astronomers because of their
simplicity and the clear predictions they make for the SEDs of T Tauri stars
and Herbig Ae/Be stars. Moreover, they can be computed quickly, which is a
great advantage when fitting observations using automated procedures. In
this paper we wish to assess the accuracy and reliability of 2-layer models,
by comparing them to detailed vertical structure models with accurate 1+1D
radiative transfer. We focus on the shape of the SED, and the predicted
height and ``flaring index'' of the disk. We first consider models where
scattering is set to zero. We find that 2-layer models overestimate
significantly the near-infrared flux, and we suggest a simple way of
correcting this effect, at least in part. At longer wavelengths, the SED of
two-layer models often show a two-bump structure, which is absent in 1+1D
models.  Nevertheless, overall agreement is reasonably good, and the
differences are in most cases within 30\%. At (sub)-mm
wavelengths the differences may even be less. The shape of the disk, as
measured by its pressure and surface scale height and by the flaring angle
are also well reproduced by two-layer models. When scattering is
included in the 1+1D models, the differences become larger, especially in
the near-infrared. We suggest simple ways to include scattering in two-layer
models and discuss their reliability. We do not compare the two-layer
models to full 2-D/3-D models, so the conclusions remain valid only within
the annulus-by-annulus approximation.}

\maketitle


\section{Introduction}
Models of disks around pre-main--sequence stars are currently used to make
predictions of a number of observable quantities and to compare them to
observations. The complexity of such models varies from so-called power-law
disks, where all the relevant disk properties are described by power-law of
radius and changed independently (see, for example, Beckwith et
al.~\citeyear{becksarg:1990}), to models where the thermal and geometrical
properties are computed self-consistently, under the assumption of
hydrostatic equilibrium (D'Alessio et al.~\citeyear{dalessiocanto:1998};
Malbet et al.~\citeyear{malbetbertout:1991},\citeyear{malbetlachaume:2001};
Bell et al.~\citeyear{bellcassklhen:1997}, \citeyear{bell:1999}; Dullemond
et al.~\citeyear{dulvzadnat:2002}). A large effort is currently being made
in improving the treatment of radiation transfer, by developing 2D codes
that deal efficiently with the very large optical depths that characterize
such disks (e.g.~Bjorkman \& Wood \citeyear{bjorkmanwood:2001}, Dullemond
\citeyear{dullemond:2002}). At the same time, it is clear that the need of
simple models, that can be easily used in interpreting the observations,
remains. This explain the success of the two-layers disk schematization
proposed some years ago by Chiang \& Goldreich (\citeyear{chianggold:1997};
henceforth CG97).

Because the CG97 model and its derivatives (e.g.~Chiang et
al.~\citeyear{chiangjoung:2001}; Dullemond, Dominik \& Natta
\citeyear{duldomnat:2001}; Lachaume et al.~\citeyear{lachmalbmon:2003}) are
often used in interpreting disk observations, it is important to estimate
their reliability and accuracy by comparing them to more detailed and
self-consistent model calculations.  At present, self-consistent 2-D and 3-D
models are still too cumbersome for such a comparison. However, there are a
number of  1D vertical structure models based on detailed radiative
transfer, which can be used as templates to assess the reliability of
2-layer models.

In this paper, we will make use of the model of Dullemond et
al.~\citeyear{dulvzadnat:2002} (henceforth DZN02). In this model the
vertical temperature profile was solved using a detailed plane-parallel 1-D
vertical radiative transfer procedure, which treats properly the slanted
penetration of the stellar radiation (often referred to as 1+1D
approximation).  Using the resulting temperature profile, the density
structure is re-computed after each iteration step by integration of the
equation of vertical hydrostatic equilibrium. In a follow-up paper
(Dullemond \& Natta \citeyear{dulnat:03a}) scattering of the stellar light
by the dust particles was included in the model.  For the case of zero
albedo this model is in fact identical to the DZN02 model, and we will use
DN03 as our ``standard 1+1D disk structure model'' against which we will
compare the two-layer models.

The goal of this paper is three-fold. First, and most importantly, we
compare an improved version of the CG97 model to the the 1+1D vertical
structure model of DN03. This initial comparison is done for dust opacities
in which scattering is switched off. Then we investigate how big the effects
of scattering are in comparison to the typical errors of the improved CG
model. Finally we discuss the possibility of using a simple recipe to
include scattering into the CG model.

\section{Description of the models}
In this section we give a short description of the improved CG97 model we
adopt, and of the 1+1D vertical structure model. Both families of
models neglect viscosity, and treat the radial dependence of the surface
density as a free parameter. Consequently, they also do not include viscous
heating which, depending on the value of the viscosity parameter $\alpha$,
could be of importance in the inner parts of the disk.

\subsection{Improved CG97 model ("\CGPLUS{}")}
The central element of the two-layer Chiang and Goldreich model is that, due
to the very shallow incident angle $\alpha\ll 1$, the impinging stellar
radiation is absorbed entirely in the very tenuous upper layers of the disk,
high above the disk's photosphere. While this layer has an optical depth of
unity for stellar radiation at that grazing angle (typically around
$\alpha=0.05$), it is very optically thin in the vertical direction
($\tau_{\mathrm{vertical}}\simeq \alpha$). Energy conservation requires that
this absorbed radiation is re-emitted as infrared radiation, half of which
is emitted away from the disk, while the other  half is emitted towards the disk
midplane. This downwards emitted flux is absorbed by the disk interior and
re-emitted once more in the infrared, though this time at longer
wavelengths. This typically gives two components in the SED: a warm
optically thin component with dust emission features in the near- to mid
infrared and a cool thermal blackbody component from the disk interior at
longer wavelengths. If the disk is optically thick to the radiation from the
midplane, then both components are equally strong, as the cool component is
a reprocessed version of the downward radiated surface emission. The
temperature $T_s$ of the surface layer is fixed: it is the optically thin
dust temperature. The emissivity of the surface layer is determined by the
surface density of the surface layer, which is calculated from energy
conservation: the total emission should equal the total absorbed flux. The
emission from the interior is determined by the temperature of the disk
interior and the disk's surface density. In the simple case of a grey
opacity and high optical depth this so-called midplane
temperature is determined by equating
the thermally emitted flux $F=\sigma T_i^4$ to the downwards directed
surface flux.  In general, $T_i$ is computed using mean opacities both for
the heating and for the cooling radiation.

The original paper of Chiang \& Goldreich (\citeyear{chianggold:1997})
describes the two-layer flaring disk model in stages. At first, a grey
opacity constant surface density model is presented for which the flaring
index $d\log(H_s)/d\log(R)=9/7$, where $H_s$ is the height of the
surface. Then a set of improvements are listed in order to account for
effects of low optical depth at various wavelength regimes. In a later paper
(Chiang et al.~\citeyear{chiangjoung:2001}) the model was refined to include
more realistic opacities and, more importantly, the self-consistent (and
numerically stable) determination of the flaring angle, which is crucial for
energy conservation.  Dullemond, Dominik \& Natta
(\citeyear{duldomnat:2001}, henceforth DDN01) subsequently added the
emission and structure of the disk's inner rim, and included effects of
self-irradiation and self-shadowing. The complete set of model equations is
listed in DDN01. In the rest of this paper this improved Chiang \& Goldreich
model will be referred to as the \CGPLUS{} model. We will not address here
issues related to the inner rim.

\subsection{Vertical structure models}
The reference model against which we compare the \CGPLUS{} model is the 1+1D
vertical structure model of DZN02 and its improved version DN03. The
radiative transfer is done in a two-stage procedure: first the primary
stellar photons are followed as they penetrate the surface layers of the
disk and get absorbed. Then the re-emitted infrared radiation is solved
using a 1-D vertical frequency and angle-dependent radiative transfer code
applied at each radial annulus. In a third stage the vertical pressure
balance is solved, under the assumption that the gas temperature equals the
dust temperature and that the gas-to-dust ratio is constant. This
three-stage procedure is then iterated until a converged solution is reached
for the complete temperature and density structure of the disk. During each
iteration, the flaring index has to be recomputed in order to compute
the flaring angle (``grazing angle'' in the language of Chiang \& Goldreich)
self-consistently. This is done by estimating the surface height of each
annulus by finding the height $H_s$ above the midplane where the direct
stellar radiation is extincted by $\exp(-1)$. The double-logarithmic
derivative of this $H_s$ with cylindric radius R is then the flaring index
from which the flaring angle can be computed (see DZN02).

Once the iteration procedure has converged, we have obtained a solution for
the dust temperature $T_{\mathrm{dust}}(R,Z)$ and the disk's gas+dust
density $\rho(R,Z)$. 

DN03 included scattering in the following way. Stage 1 of the DZN02 model
(the irradiation of the disk by primary stellar photons) was replaced with a
1-D Monte-Carlo code. Primary stellar photons are followed as they penetrate
into the surface of the disk and scatter around until they either get
absorbed, or escape to infinity.  If a photon is absorbed, it leaves behind
its energy, which is then used as a source term for the radiative transfer
of the second stage. In the second stage, which takes care of the re-emitted
infrared radiation from the disk, scattering is not included. This is a
valid approximation if the dust grains are small enough (typically not
larger than 0.5 $\mu$m).

\section{Comparing models without scattering}
As a first step we compare the \CGPLUS{} model and the 1+1D vertical
structure for a test case without scattering. In this way the physical
assumptions of both models are the same, so that in principle we expect both
models to produce very similar SEDs. The comparison is first done for a
single annulus at various radii and optical depths, irradiated by a T Tauri
star or by a Herbig Ae/Be star. We then compare the SEDs of models of entire
disks, as well as intensity profiles at some selected wavelengths.

\subsection{Single annulus}\label{subsec-annulus-setup}
As a first example we consider an annulus of a disk at 1 AU from the
central star, with width  0.01 AU (i.e.~an annulus between 1.00 and 1.01
AU). We take the central star to be a Herbig Ae/Be star ($M_{*}=2M_{\odot}$,
$R_{*}=2R_{\odot}$, $T_{\mathrm{eff}}=10000$K). To simplify matters, we
assume the spectrum of the star to be a blackbody spectrum with
$T=T_{\mathrm{eff}}$. The grazing angle (i.e.~the angle at which the stellar
radiation enters the disk's atmosphere) is computed self-consistently in
both \CGPLUS\ and 1+1D models, but for the purpose of the comparison
described in this section we assume it to be fixed to $\alpha=0.05$.  We
also assume, for these annulus tests, that half of the surface of the star
as seen from the top of the annulus is covered by the very inner parts of
the disk (see CG97). This effectively reduces the stellar flux by a factor
0.5. The absorption cross section is that of a silicate grain of 0.1 $\mu$m
(Draine \& Lee \citeyear{drainelee:1984}), with zero albedo. 
Different choices of the opacity will not change the results of this
paper in any significant way, unless explicitely noted. The surface
density of the disk is such that the total vertical absorption optical depth
(from $z=-\infty$ to $z=\infty$) at 550 nm is $\tau_{550nm}=10$. The mean
molecular weight is assumed to be $\mu=2.3$.

\begin{figure}
\centerline{
\includegraphics[width=9cm]{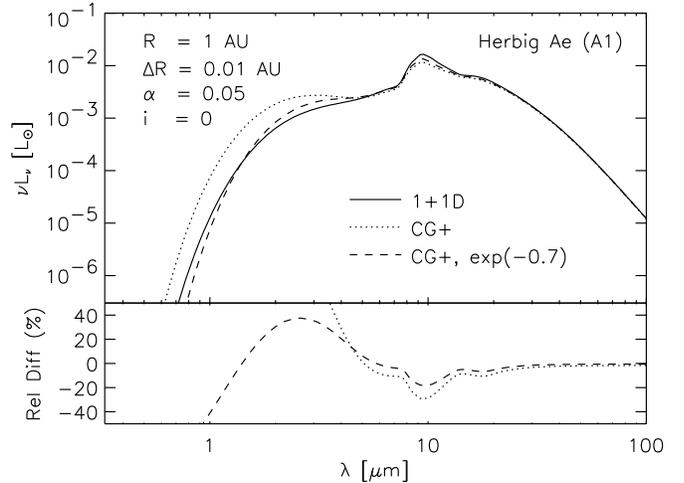}}
\caption{The SED for the single-annulus test case nr.~A1 (Herbig Ae star) at
1 AU, compared to the \CGPLUS{} model in its two proposed
variants.}
\label{fig-haebe-annul-sed}
\end{figure}

In Fig.~\ref{fig-haebe-annul-sed} the SED of this annulus is shown for both
the vertical structure model and the \CGPLUS{} model. The difference between
the models is less than 30\% at wavelengths $\lambda>8 \mu$m. 
At shorter wavelengths the \CGPLUS{} model has a bump that is not present in the
SED of the vertical structure model. The reason can be traced back to the
surface temperature assumed in the \CGPLUS{} model, which is that 
of grains
in an optically thin medium. In reality, however, only a fraction of the
grains in the surface layer see the unextincted stellar light. More than 36
\% of the material in the surface layer lies at optical depths  larger
than unity (along the grazing incident path of the stellar radiation).  It
is hard to assign a single temperature to the material in the surface
layer. But a reasonable average temperature could be the temperature at an
optical depth of $\tau=0.7$, which is the location where exactly half of the
stellar radiation has been absorbed. The dashed line in
Fig.~\ref{fig-haebe-annul-sed} shows the SED as computed by the \CGPLUS{}
model in which the surface temperature is evaluated at an extinction of
$\exp(-0.7)=0.5$.

\begin{figure}
\centerline{
\includegraphics[width=9cm]{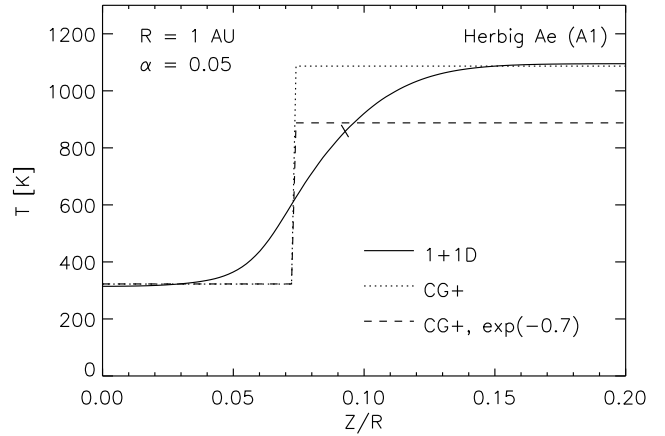}}
\caption{The temperature structure for the single-annulus test case A1
(Herbig Ae star), compared to the \CGPLUS{} model in its two proposed
variants. The transition from the surface to the midplane temperature in the
\CGPLUS\ models occurs at $H_s$, i.e., where the optical depth along a ray
at a grazing angle $\alpha=0.05$ at stellar wavelengths is $\tau_\ast \sim
1$. The tick mark on the solid curve shows the location of $H_s$ in the 1+1D
case.}
\label{fig-haebe-annul-vertstr}
\end{figure}

In Fig.~\ref{fig-haebe-annul-vertstr} the temperature structure of the same
annulus is shown as a function of the vertical coordinate $Z/R$, where $Z$
is the height above the midplane, and $R$ the radius from the star (in this
case $R=1\AU$). High above the surface of the disk the temperature is
constant, and virtually equal to the optically thin dust grain temperature
at that distance from the star. The slight difference is due to the fact
that in the 1-D model the dust high above the disk is heated not only by the
direct stellar radiation but also by the thermal emission of the disk below
it.

As one goes towards smaller $Z$ the temperature starts to decrease due to
extinction of the direct stellar flux by the material of the disk. This is
the location of the disk's ``hot surface layer'' which produces the dust
emission features. The small tick mark on the temperature curve denotes the
surface height $H_s$ where the grazing optical depth is unity, i.e.~where
most of the direct stellar radiation is absorbed. Continuing downwards from
the surface layers, the temperature slope levels off again because the
disk's own infrared radiation becomes competitive with the direct stellar
radiation in the thermal balance equation of the dust grains. This is where
the disk's photosphere is located and where most of the thermal continuum
emission is produced (the midplane emission, in the terminology of CG
models).

Over-plotted over the smooth temperature curve is the  two-layer
\CGPLUS{} model, in its two variants mentioned above. The sharp step in the
temperature profile is where the surface layer of the \CGPLUS{} model
starts: $z=H_s$. The main difference in the two variants of the model is the
temperature of this surface layer. For one variant, the surface layer
temperature equals that of optically thin dust grains. For the corrected
variant this temperature is lower, and is closer to the temperature of the
vertical structure model at the location of the tick mark.

Note that $H_s$ is different in \CGPLUS\ and 1+1D models. Most of
difference is due to the fact that in  the 1+1D 
models the density in the disk surface deviates from the Gaussian shape
predicted by isothermal models (see, for example, Fig.~3 of DN03). 
An additional, smaller
difference is due to the fact that \CGPLUS\ models use
the mean opacity averaged over the stellar flux
to define
$H_s$, while the 1+1D models use the wavelength-dependent opacity.

\begin{table}
\centerline{
\begin{tabular}{c|ccc|cc}
  & $M_{*}$ & $R_{*}$  & $T_{*}$ & 
$R$ & $\tau_{V}$\\
\hline
A1 & 2   & 2 & 10000 & 1.  & 10 \\
A2 & 0.5 & 2 &  4000 & 0.1 & 10 \\
A3 & 2   & 2 & 10000 & 50. & 300 \\
\end{tabular}}
\caption{\label{table-params}
The parameters of the single-annulus models used in the comparison.
The first column is the identification used in this paper. 
Columns 2-4 are the stellar parameters (in units of $M_{\odot}$, $R_{\odot}$
and K) and columns 5-6 are the annulus parameters. The radius is in AU.
All model annuli have a width $\Delta R/R=0.01$ and a grazing incident angle
of stellar radiation of $\alpha=0.05$.}
\end{table}

\begin{figure}
\centerline{
\includegraphics[width=9cm]{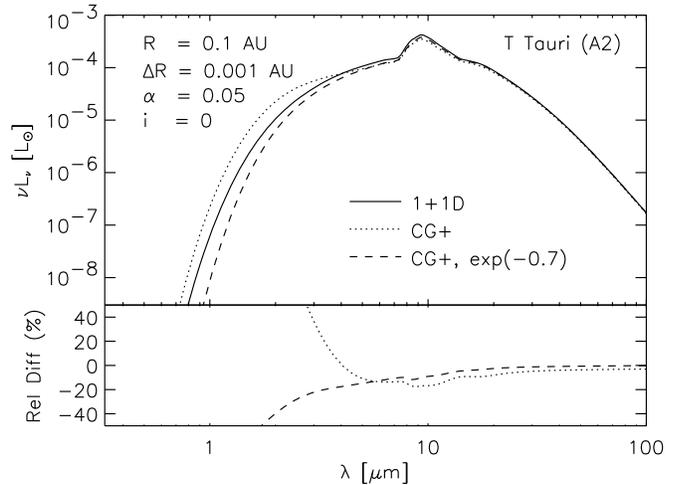}}
\caption{The SED for the single-annulus test case nr.~A1 (Herbig Ae star) at
1 AU, compared to the \CGPLUS{} model in its two proposed
variants.}
\label{fig-ttauri-annul-sed}
\end{figure}

Table \ref{table-params} lists the parameters of two other annuli.  Test A2
is the T Tauri analog of A1, with lower stellar mass and temperature, and
the annulus taken at a smaller radius (to obtain similar temperatures in the
disk). In Fig.~\ref{fig-ttauri-annul-sed} the SED of this test case is
shown. The results are qualitatively similar to the case of the Herbig Ae
star (test A1). In this particular example, however, the $\exp(-0.7)$
correction of the surface temperature calculation in the \CGPLUS\ model does
not make a major improvement, although also in this case the overall SED of
the 1+1D model is slightly better reproduced when this factor is taken into
account.

\begin{figure}
\centerline{
\includegraphics[width=9cm]{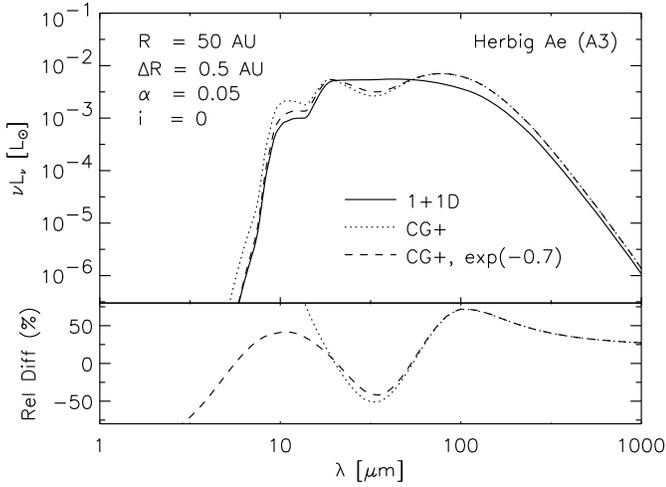}}
\caption{The SED for the single-annulus test case nr.~A3 (Herbig Ae star,
high optical depth) at 50 AU, compared to the \CGPLUS{} model
in its two proposed variants.}
\label{fig-haebe-annul-thick-sed}
\end{figure}

\begin{figure}
\centerline{
\includegraphics[width=9cm]{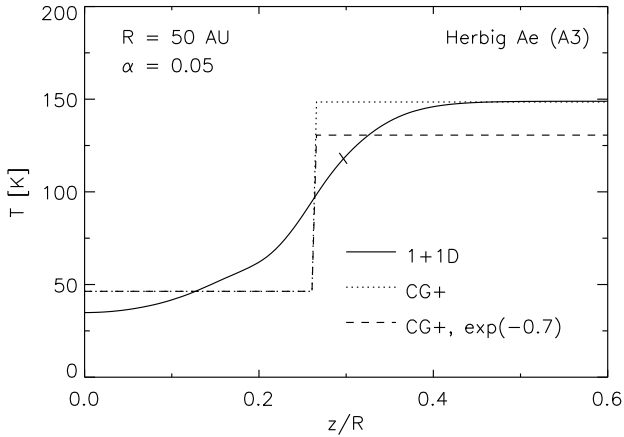}}
\caption{The temperature structure form model A3.}
\label{fig-haebe-annul-thick-temp}
\end{figure}

As shown above, in many cases the \CGPLUS{} model is reasonably well in
agreement with the 1+1D vertical structure model. Yet, there are cases in
which the differences are appreciable. We show an example in
Fig.~\ref{fig-haebe-annul-thick-sed} which plots the SED of the A3 annulus.
This annulus has $\tau_V=300$, and is at 50 AU from our Herbig Ae star.  It
is clearly seen that in this case the difference between \CGPLUS{} and 1+1D
predictions is rather large, though still within a factor of 2. The
explanation for the breakdown of the \CGPLUS{} model in this case is likely
the use of mean opacities to compute the midplane temperature.  As noted by
DZN02, the midplane of the disk can cool down by leaking radiation at long
wavelengths (where the optical depth of the disk becomes smaller than
unity), more than expected on the basis of a mean-opacity analysis. This is
clearly seen in Fig.~\ref{fig-haebe-annul-thick-temp}, which shows the
temperature structure of the A3 annulus. The 1+1D temperature profile, which
goes from values lower than predicted by \CGPLUS\ in the midplane to larger
ones at intermediate $z$ results in lower fluxes around 100 $\mu$m and
higher fluxes in the mid-infrared, so that the two-bump shape of \CGPLUS\
practically disappears in the 1+1D model.

\subsection{Full disk models}
In this section we compare the SED and the physical
structure of entire disks
computed with the 1+1D vertical structure model and \CGPLUS{} models. We
present three illustrative cases: a disk around a Herbig Ae star, a T Tauri
star and a Brown Dwarf. Table \ref{table-params-fulldisk} lists the
parameters of the three disk models.

\begin{table}
\centerline{
\begin{tabular}{c|ccc|ccc|c}
 & $M_{*}$ & $R_{*}$  & $T_{*}$ & 
$R_{\mathrm{in}}$  &$R_{\mathrm{out}}$  & $\Sigma(1AU)$ & $M_{\mathrm{disk}}$ \\
\hline
F1 & 2   & 2   & 10000 & 1    & 300 & 1000 & 0.2   \\
F2 & 0.5 & 2   &  4000 & 0.1  & 300 & 100  & 0.02  \\
F3 & 0.1 & 1.3 &  2600 & 0.033 & 30  & 100  & 0.002 \\
\end{tabular}}
\caption{\label{table-params-fulldisk} The parameters of the full disk
models. The first column is the identification used in this paper. 
Columns 2-4 are the stellar parameters (in units of $M_{\odot}$,
$R_{\odot}$ and K) and columns 5-6 are the annulus parameters: outer radius
in AU, and the surface density $\Sigma$ at 1 AU. All models have 
a power law index for the surface density of
-1 ($\Sigma\propto 1/R$). Column 8 is the mass of the disk in units of 
$M_{\odot}$ as derived from the other parameters.}
\end{table}

We assume that the disk is in hydrostatic equilibrium and compute the
flaring angle is self-consistently. This is done by evaluating the flaring
index
\begin{equation}
\xi \equiv \frac{d\lg(H_s/R)}{d\lg R}
\comma
\end{equation}
where $H_s$ is the height of the $\tau_{\mathrm{grazing}}=1$ surface of the
disk. From the flaring index, the flaring angle can be computed directly:
\begin{equation}
\alpha = 0.4 \frac{R_{*}}{R} + \xi \frac{H_s}{R}
\fullstop
\end{equation}
For numerical stability reasons this flaring index is always evaluated 2
grid points inwards of the point where it is used (see Chiang et
al.~\citeyear{chiangjoung:2001}).  In these models we assume that the 
stellar flux is not reduced due to occultation of the central star
by the inner disk,
as was assumed in the single-annulus models.

\begin{figure}
\centerline{
\includegraphics[width=9cm]{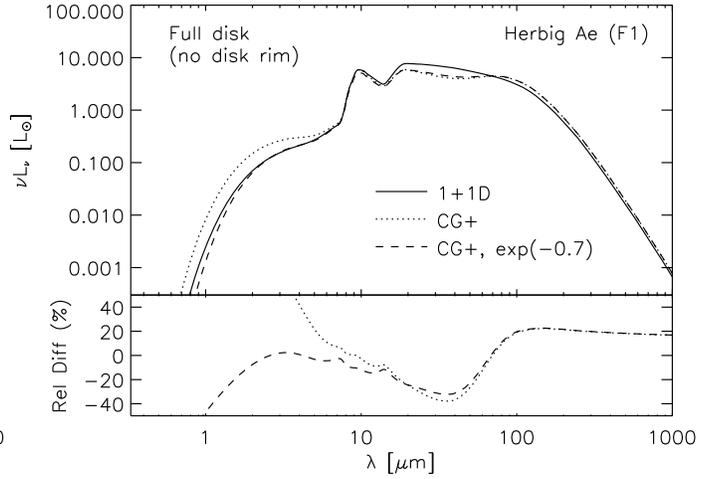}}
\caption{The SED for the full disk model F1 (Herbig Ae star), 
compared to the \CGPLUS{} model in its two proposed variants.}
\label{fig-haebe-fulldisk-sed}
\end{figure}

Fig.~\ref{fig-haebe-fulldisk-sed} shows the SED of model F1.
Over-plotted are the two proposed \CGPLUS{} models (the one with
optically thin surface temperature and the one with the correction factor
$\exp(-0.7)$). One can see that overall the models agree, but that
the \CGPLUS{} model clearly shows a double-bump structure between 20 $\mu$m
and 100 $\mu$m which is not seen in the full vertical structure model. 
\begin{figure}
\centerline{
\includegraphics[width=9cm]{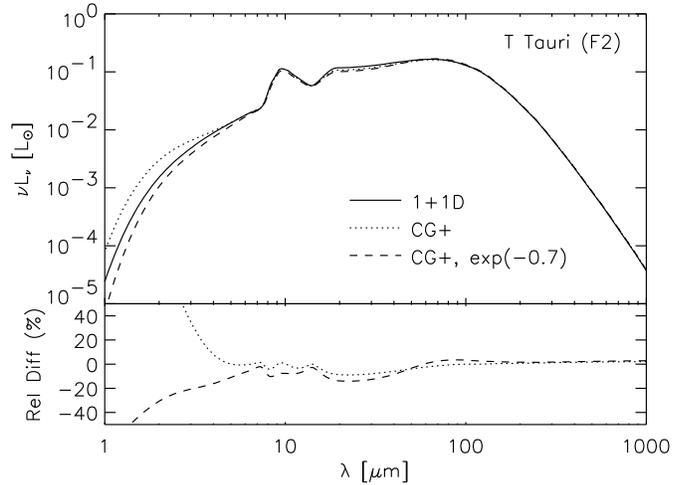}}
\caption{Same as Fig.~\ref{fig-haebe-fulldisk-sed}, but now for model
F2 (a T Tauri star).}
\label{fig-tt-fulldisk-sed}
\end{figure}
\begin{figure}
\centerline{
\includegraphics[width=9cm]{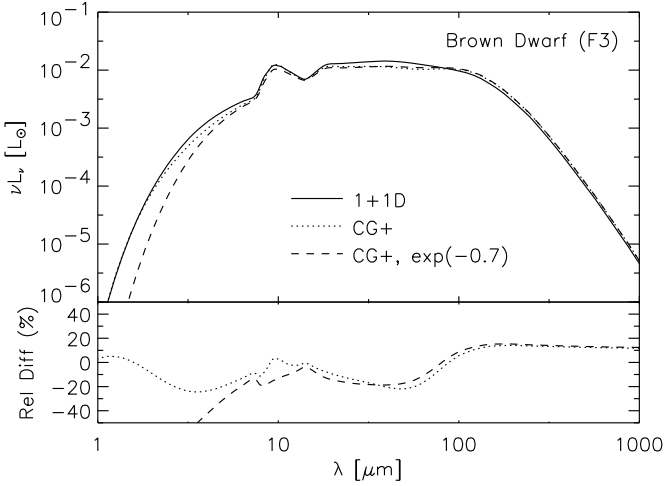}}
\caption{Same as Fig.~\ref{fig-haebe-fulldisk-sed}, but now for model
F3 (a Brown Dwarf).}
\label{fig-bd-fulldisk-sed}
\end{figure}
In Fig.~\ref{fig-tt-fulldisk-sed} the same is plotted, but now for a T Tauri
star (model F2). The disk is taken to be less massive than the Herbig star
disk. The \CGPLUS{} model works really well for this case, with errors
mostly within 10\%. Finally, in Fig.~\ref{fig-bd-fulldisk-sed} the SED of a
Brown Dwarf disk model is show (model F3). This disk is taken to have a much
smaller outer radius and mass than the T Tauri case. Nevertheless,
the disk remains very optically thick, as in the other models. 
Again, errors are
reasonably small, though slightly larger than for model F2.

\begin{figure}
\centerline{
\includegraphics[width=9cm]{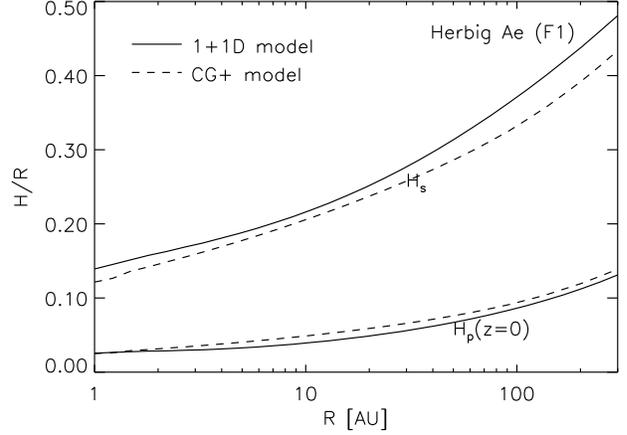}}
\centerline{
\includegraphics[width=9cm]{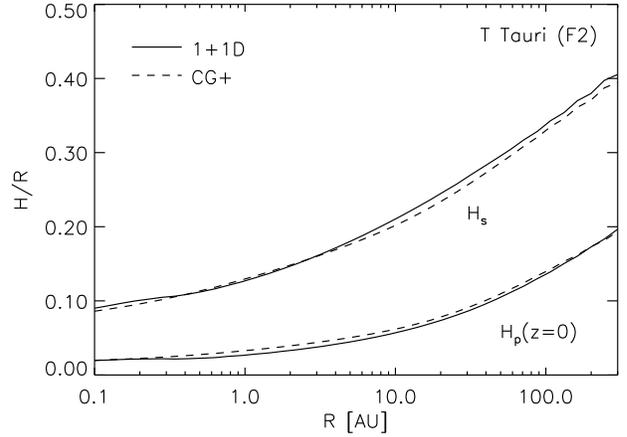}}
\caption{Top panel: Pressure scale height $H_p$
and  surface height $H_s$ as function of $R$ for
model F1 (Herbig Ae star). The solid line shows the results
of the vertical structure model, the dashed line those of
\CGPLUS{} ones. Bottom Panel: same for model F2 (T Tauri star).}
\label{fig-tt-fulldisk-hphs}
\end{figure}

In Fig.~\ref{fig-tt-fulldisk-hphs} the height of the disk of model F1
(Herbig Ae star) and F2 (the T Tauri case) is shown as a function of radius,
for both the 1+1D model and the \CGPLUS{} model.  The pressure scale height
is derived from the temperature of the disk at the equatorial plane. At very
small radii $H_p$ is virtually the same in \CGPLUS{} and 1+1D models.  At
larger radii, $H_p$ in the 1+1D model drops below that of the \CGPLUS{}
model, because the equatorial plane temperature of the 1+1D model tends to
be lower, as discussed in Section \ref{subsec-annulus-setup}.  The
difference, however, is always very small.

The surface scale height in $H_s$ tends to be  larger
in 1+1D models, reflecting in part the deviation of the very upper layers
from a Gaussian profile,  as
discussed of Section  \ref{subsec-annulus-setup}. The degree by which \CGPLUS\ models can reproduce
the results of 1+1D ones depends on the dust opacity in combination with 
other model parameters, such as the disk mass. 
Here we just note that the difference is larger for the HAe star,
and for larger radii.
However,
these differences remain  small and we can conclude that, in all cases we
have considered, the disk shape can be derived with good accuracy using the
simple approximations of the \CGPLUS\ models.

\begin{figure}
\centerline{
\includegraphics[width=9cm]{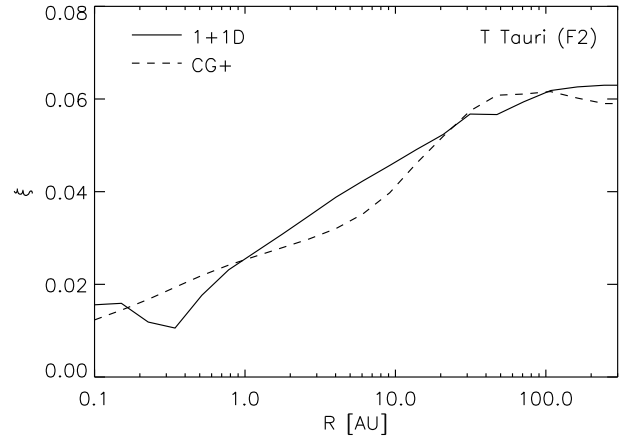}}
\caption{The flaring index of the disk $\xi$ is plotted as a function of
radius for model F2 (T Tauri star).  The vertical structure model is shown
as a solid line, while the \CGPLUS{} is the dashed line.}
\label{fig-tt-fulldisk-flareidx}
\end{figure}

Fig.~\ref{fig-tt-fulldisk-flareidx} shows the flaring index as function of
radius for model F2. This quantity gives information on the amount of
radiation emitted by the disk at any given radius, because for $R\gg R_{*}$
the emission is proportional to $\xi$. Although the run of $\xi$ with $R$ is
similar in the two models, we note that $\xi$ is a smooth function of radius
in the \CGPLUS\ models, but not in the 1+1D one. In particular, the kink in
the flaring index 0.3 AU is not due to numerical inaccuracies but rather to
the behavior of the midplane optical depth , which in this disk model drops
below unity at long wavelengths at about that radius. The consequent cooling
of the midplane below optically thick temperatures, discussed in \S 3.1, is
seen as a slight kink in $H_s$ in Fig.~\ref{fig-tt-fulldisk-hphs}.  This
effect, which depends on the disk and stellar parameters and on the exact
wavelength dependence of the dust opacity, can have important consequences
on the stability of 1+1D models.

\begin{figure}
\centerline{
\includegraphics[width=9cm]{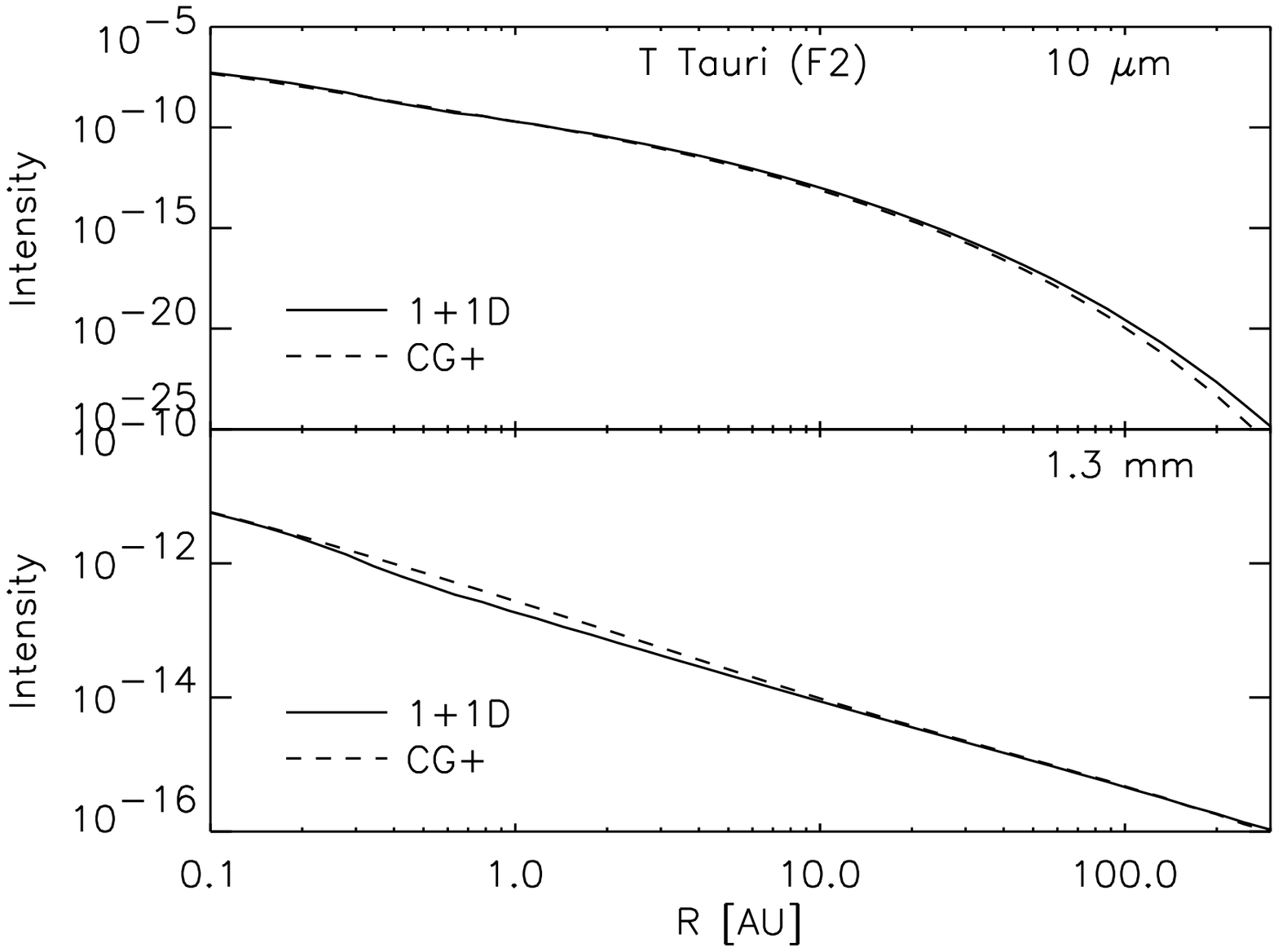}}
\caption{The brightness profile of model F2 (T Tauri star) as a function of
radius at two different wavelengths.  The 1+1D model results are shown as a
solid line, the \CGPLUS{} model results as a dashed line.}
\label{fig-tt-bright}
\end{figure}

Fig.~\ref{fig-tt-bright} shows the brightness profiles of the two model
approaches at two wavelengths, 10 $\mu$m and 1.3mm, where current
instrumentation has the spatial resolution required to resolve the disks.
The differences between the models are very small, probably within the
observational accuracy of most experiments.  The 10 $\mu$m profile has a
clear exponential cut-off at large radii due to the fact that the dust
becomes too cool to emit strongly at mid-infrared wavelengths.  The
\CGPLUS{} and 1+1D models produce virtually identical emission at this
wavelength, as seen also in the SED (Fig.~\ref{fig-tt-fulldisk-sed}).  At
1.3 mm the differences are slightly larger (up to 30\%), which reflects the
fact that sometimes the 1+1D models can be slightly cooler in the disk's
midplane than the \CGPLUS{} models, as discussed above.

\section{Models with scattering}

\subsection{Effect of scattering on the disk}
The effects of scattering of the
stellar radiation on the structure and SED of  circumstellar disks
was discussed in detail in DN03. The main effect is that the thermal
infrared flux from the disk is reduced at all wavelengths, with a
larger effect at shorter than at longer
wavelengths. 
\begin{figure}
\centerline{
\includegraphics[width=9cm]{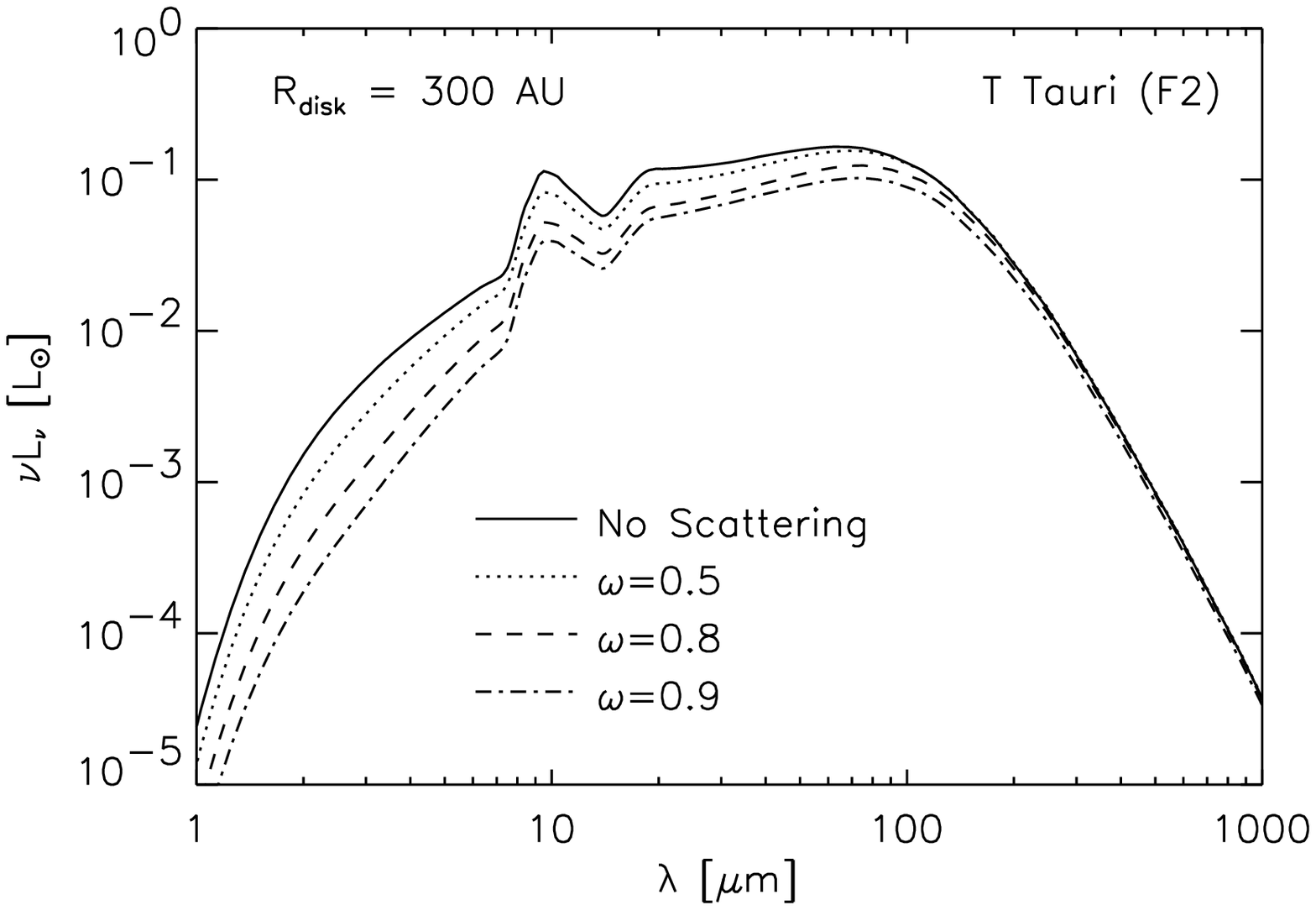}}
\caption{Example of the effect of scattering on the SED of a T Tauri disk
(model F3). Only the thermal emission of the disk is shown in this
figure. Scattered light would appear as a strong rise towards short
wavelengths.}
\label{fig-example-sed-redux}
\end{figure}
This reduction takes place because the scattering reflects away from the
disk part of the stellar radiation, which therefore cannot be reprocessed
into infrared radiation. The reflected fraction of the incident flux can be
evaluated analytically using the H-functions of Chandrasekhar, which were
discussed in this context in DN03. For a moderate value of the albedo
($\omega\sim 0.5$) and isotropic scattering the reflected fraction is about
25\% (of order $\omega/2$), and the reduction factor of the disk's thermal
emission is about 0.75, when averaged over the infrared spectrum. For higher
albedo the reduction is larger, while it decreases if the scattering is more
forward-peaked.

The reason why the reduction is stronger at short wavelength than at long
wavelengths can be traced back to the fact the surface layers of the disk
are affected more strongly by  scattering than the interior. A photon
that gets lost in the disk's interior may scatter many times and still get
absorbed rather than escape. Therefore the deep interior layers of the disk
will have a temperature only slightly below the temperature of a disk
without scattering.

The SEDs of a T Tauri disk (model F2) with increasing albedo,
computed as in DN03,  are
shown in Fig.~\ref{fig-example-sed-redux}.
The effect of reddening of the SED as $\omega$ increases is clearly seen.
This
suggests that \CGPLUS\ model-predicted SEDs, which tend in general 
to overestimate the amount of flux at shorter wavelengths with respect
to 1+1D models, may be  significantly in error when realistic
dust models are considered.

\subsection{A simple recipe for scattering in \CGPLUS\ models: is it possible?}

Although the qualitative effect of scattering can be easily understood ``a
priori", it is more difficult to estimate quantitatively how the infrared
spectrum of the disk is affected at each individual wavelength. Still, it
is important to explore the possibility of including scattering in the
\CGPLUS\ models in some simple way, and to assess the reliability of any
such model by comparing its predictions to those of the more accurate 1+1D
models.

A first recipe is based on the following considerations. In \CGPLUS\ models,
the stellar radiation is intercepted by the disk surface, which will re-emit
(part as scattered light and part as reprocessed radiation) 1/2 of it toward
the observer and 1/2 toward the midplane.  This entire second 1/2 will
be absorbed by the midplane, whose temperature, therefore, will not depend
on $\omega$. Of the 1/2 emitted toward the observer, a fraction $\omega$ is
in the form of scattered light, and consequently only a fraction
(1-$\omega$) is in the form of thermal reprocessed flux at infrared
wavelength. Since the temperature of this surface layer dust depends
only on its absorption cross section, one can compute the SED by combining
the emission of the midplane of a \CGPLUS\ model with $\omega=0$ with the
surface emission reduced by a factor (1-$\omega$/2).  This recipe neglects
multiple scattering and any wavelength-dependent effect. However, for low
albedo it reproduces well the overall reduction of the infrared excess, and
it is based on very simple arguments.

A second possibility, suggested also by CG97, is to apply the classical
result that the fraction of incident photons absorbed by a plane-parallel
slab is reduced by a factor $\eta=\sqrt{1-\omega}$. The midplane will then
have a temperature (in the optically thick case) lower than T($\omega=0$) by
a factor $\eta^{1/4}=(1-\omega)^{1/8}$. The surface grain temperature is
again unchanged, but the infrared emission of the surface needs to be
reduced by a factor $\eta$.  For not too high albedo, also this recipe
will reproduce the overall reduction of the infrared excess found by DN03.

\begin{figure}
\centerline{
\includegraphics[width=9cm]{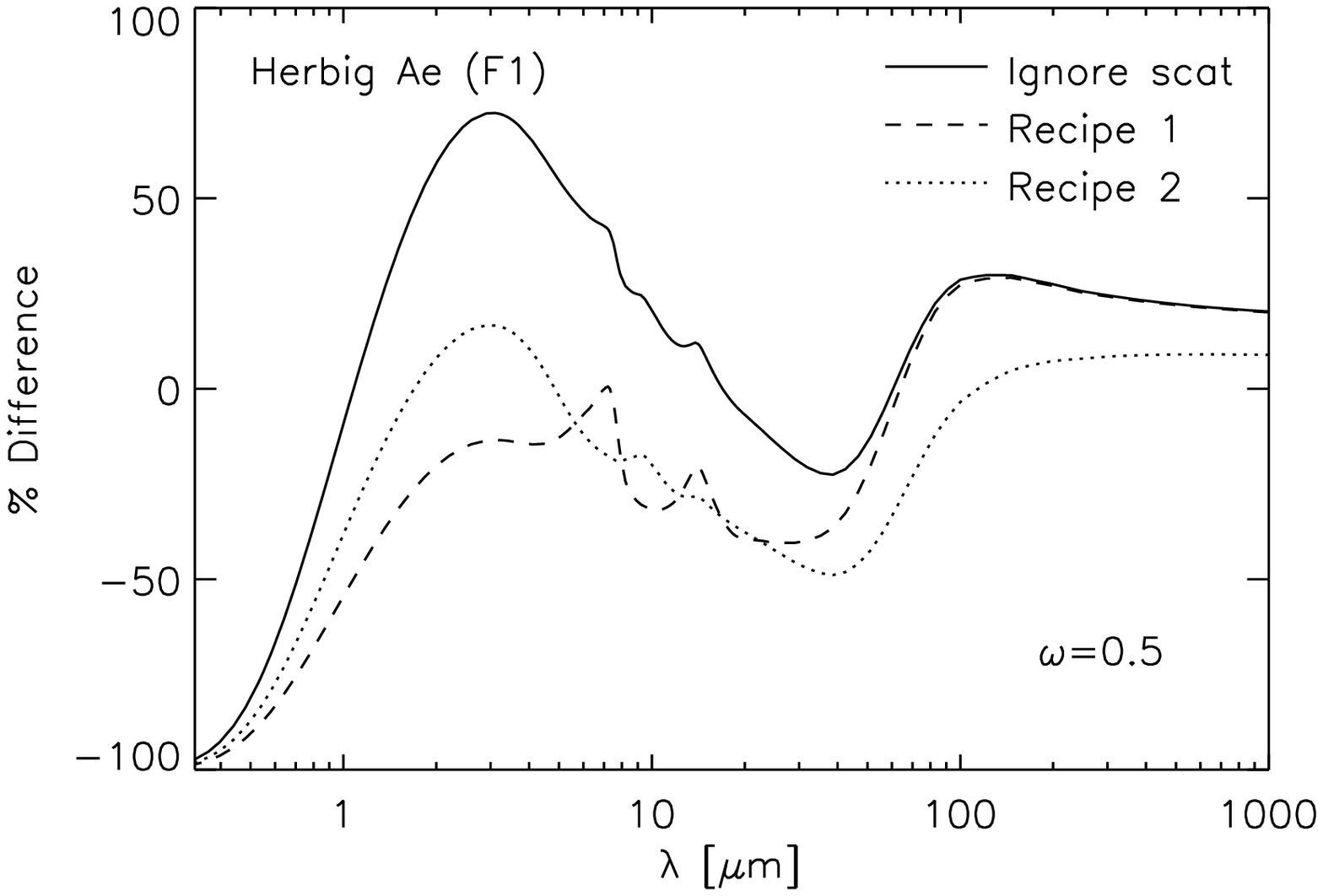}}
\centerline{
\includegraphics[width=9cm]{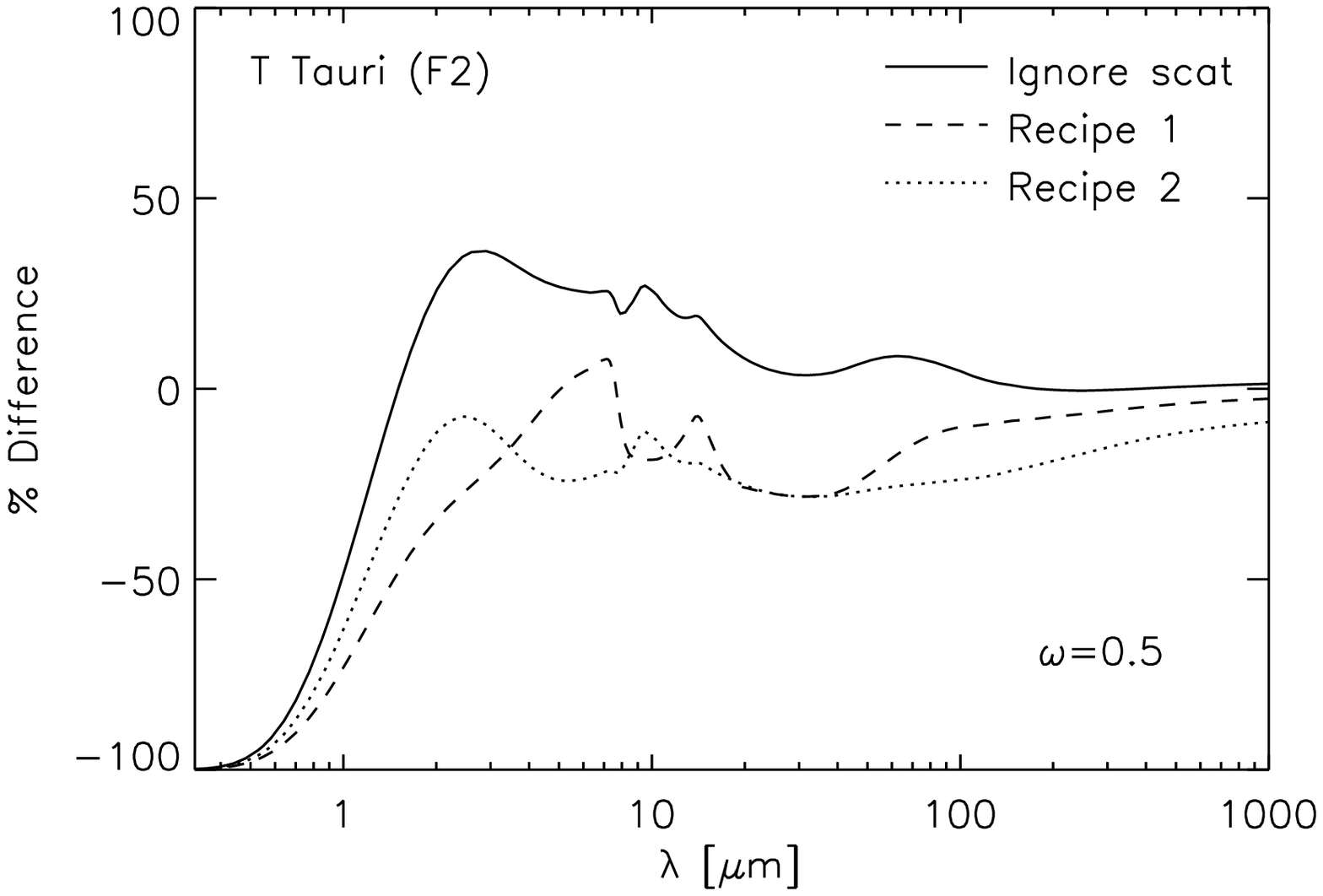}}
\caption{\label{fig-seddiff-annul-cgrecip} The factor by which scattering
suppresses the SED of the disk annulus of test cases A1 and A2 for an albedo of
$\omega=0.5$.  The solid line is the suppression factor from the 1-D
vertical structure model. The other lines are for the two recipes
proposed for the \CGPLUS{} model.}
\end{figure}

We show a comparison of these two prescriptions with the correct results of
DN03 in Fig.~\ref{fig-seddiff-annul-cgrecip}.  The top panel shows the
comparison for the Herbig Ae star disk (F1), the bottom panel for the T
Tauri star disk (F2). Each curve shows the fractional difference of the SED
(reprocessed radiation only) of \CGPLUS\ models, as labeled, to that of the
comparison model, which is the 1+1D one with $\omega=0.5$ (isotropic
scattering).  The solid line shows the results for \CGPLUS\ models where
scattering is ignored. One can see that these models overestimate the flux
at all wavelengths, and that the discrepancy can be larger than a factor of
two (for the Herbig Ae disk) between about 2 and 5 $\mu$m.  At longer
wavelengths the agreement improves, and becomes quite good below about 100
$\mu$m.

The dashed and dotted lines show the results obtained with the \CGPLUS{}
model using recipe 1 and recipe 2, respectively.  Both prescriptions seem to
work equally well, and improve the SED description at wavelengths longer
than about 2--3 $\mu$m, reducing the discrepancies below $\sim$ 20\%. Note
that the larger difference in the Herbig Ae SED around 40 $\mu$m is not
related to scattering (see \S 3.2).  Although we do not find any strong
reason to prefer recipe 1 over recipe 2, we note that recipe 2 reproduces
more correctly the overall reduction of the infrared excess for large values
of $\omega$ (compare the run of $\eta$ with $\omega$ with the results of
Fig.1 in DN03).  Finally, it is interesting to notice that recipe 1 tends to
suppress the 10 $\mu$m feature with respect to the continuum, while this is
not the case for recipe 2.

\section{Summary and conclusion}

In this paper we analyze the reliability of two-layer models for passive
irradiated flaring dusty disks around pre-main-sequence stars by comparing
their SEDs with those computed from more realistic 1+1D vertical structure
models. The two-layer model used in this paper is called \CGPLUS{}, and is
an improved version of the Chiang \& Goldreich model (see DDN01).  The 1+1D
model involves detailed 1-D vertical radiative transfer which treats
scattering of the primary stellar photons penetrating the surface layers of
the disk (see DN03). All the models presented in this paper can be
downloaded as ASCII tables from {\tt http://www.mpa-garching.mpg.de/
PUBLICATIONS/DATA/radtrans/cgcompare/}.

We start with a comparison of the SEDs for different disk models where
scattering is ignored. In general, \CGPLUS\ models over-predict the emission
at short wavelengths. This is caused by the assumption that the surface
grains have the temperature of grains in an optically thin medium. In many
cases one can obtain a better agreement by computing the surface temperature
with at an attenuation exp(-0.7) of the stellar flux.  At longer wavelengths
(mid and far-infrared), the \CGPLUS{} SEDs sometimes shows two maxima,
due to the two components of the model: one at mid-infrared wavelength
(surface layer) and one at far infrared wavelengths (interior) (see, for
example, Fig.~\ref{fig-haebe-fulldisk-sed}). The 1+1D model, on the other
hand, always produces a smooth SED (with the exception of dust
features), which looks qualitatively somewhat different. However, a closer
look shows that the agreement between different models is often quite good
(within about 20\%), with some potential problems deriving from the use of
mean opacities in \CGPLUS\ models.  In our set of models, the largest
discrepancies occur for the Herbig Ae star, where they reach $\sim$40\%
around 40 $\mu$m.  The shape and strength of the 10 micron emission feature
is in general well predicted by the \CGPLUS{} models, and so is the
peak-over-continuum ratio of this feature and its intensity profile. We
therefore argue that also other dust features will be reasonably well
reproduced by \CGPLUS{} models.  At millimeter wavelengths, both
models predict similar fluxes (within 20\% at most) and intensity profiles.

A comparison of the disk physical structure is also interesting.  Both the
pressure scale height $H_p(R)$, which depends on the midplane temperature,
and the surface height $H_s(R)$ predicted by the \CGPLUS{} model agree
reasonably well with the 1+1D models.

Note that we have used in this paper an {\it improved} version of two-layer
code, which does not make use of the analythical expressions for flaring
angle and temperatures derived in CG97, but derives the disk structure
self-consistently.  A comparison of the CG97 analythical model with the
results of 1D calculations can be found in Kraus \& Kr\"ugel (in
preparation).

\CGPLUS{} models do not include scattering of the stellar light from dust
grains.  When compared to the more accurate 1+1D models which include
scattering, even for quite low values of the albedo or rather peaked-forward
scattering, we find that the differences in the SED can be larger than a
factor of two in the near infrared. However, as the effect of scattering
itself decreases at longer wavelengths, so do the differences between
models.  We have implemented two different simple recipes to our \CGPLUS\
code to include scattering without changing the basic idea on which
two-layer models are built. Both recipes seem to improve somewhat the
agreement of \CGPLUS\ models with the results of the 1+1D calculations,
although recipe 2 is probably a better approximation for very large values
of $\omega$.  (see DN03).

In summary, the results from this paper show that \CGPLUS\ models
compare well with the much more detailed 1+1D vertical structure models.
These two-layer models can therefore be used with reasonable confidence to
fit observed SEDs of actual objects, provided that the parameters are not
stretched too far. Problems could occur, for instance, when the disks are
not optically thick enough, so that the annulus-by-annulus approach, on
which both the \CGPLUS\ and the 1+1D models are based, breaks down. 
In such a case one must resort to a full 2-D or even 3-D treatment of the
problem.  A future study should find out under which circumstances such a
2-D/3-D approach is unavoidable, or, oppositely, when the
annulus-by-annulus approach (and thereby the \CGPLUS\ approach) is
justified.

\end{document}